\documentclass[pdflatex,sn-mathphys-num]{sn-jnl}

\usepackage{graphicx}%
\usepackage{multirow}%
\usepackage{amsmath,amssymb,amsfonts}%
\usepackage{amsthm}%
\usepackage{mathrsfs}%
\usepackage[title]{appendix}%
\usepackage{xcolor}%
\usepackage{textcomp}%
\usepackage{manyfoot}%
\usepackage{booktabs}%
\usepackage{algorithm}%
\usepackage{algorithmicx}%
\usepackage{algpseudocode}%
\usepackage{listings}%

\theoremstyle{thmstyleone}%
\theoremstyle{thmstyletwo}%

\theoremstyle{thmstylethree}%

\begin{document}

\title[Article Title]{Using Information Geometry to Characterize Higher-Order Interactions in EEG}

\author*[1]{\fnm{Eric} \sur{Albers}}\email{eric.albers@uleth.ca}

\author[2]{\fnm{Paul} \sur{Marriott}}\email{pmarriott@uwaterloo.ca}

\author[1]{\fnm{Masami} \sur{Tatsuno}}\email{tatsuno@uleth.ca}

\affil*[1]{\orgdiv{Department of Neuroscience}, \orgname{University of Lethbridge}, \orgaddress{\street{4401 University Dr W}, \city{Lethbridge}, \postcode{T1K 3M4}, \state{Alberta}, \country{Canada}}}

\affil[2]{\orgdiv{Department of Statistics and Actuarial Science}, \orgname{University of Waterloo}, \orgaddress{\street{200 University Ave W}, \city{Waterloo}, \postcode{N2L 3G1}, \state{Ontario}, \country{Canada}}}

\abstract{In neuroscience, methods from information geometry (IG) have been successfully applied in the modelling of binary vectors from spike train data, using the orthogonal decomposition of the Kullback-Leibler divergence and mutual information to isolate different orders of interaction between neurons. While spike train data is well-approximated with a binary model, here we apply these IG methods to data from electroencephalography (EEG), a continuous signal requiring appropriate discretization strategies. We developed and compared three different binarization methods and used them to identify third-order interactions in an experiment involving imagined motor movements. The statistical significance of these interactions was assessed using phase-randomized surrogate data that eliminated higher-order dependencies while preserving the spectral characteristics of the original signals. We validated our approach by implementing known second- and third-order dependencies in a forward model and quantified information attenuation at different steps of the analysis. This revealed that the greatest loss in information occurred when going from the idealized binary case to enforcing these dependencies using oscillatory signals. When applied to the real EEG dataset, our analysis detected statistically significant third-order interactions during the task condition despite the relatively sparse data (45 trials per condition). This work demonstrates that IG methods can successfully extract genuine higher-order dependencies from continuous neural recordings when paired with appropriate binarization schemes.}


\maketitle

\section{Introduction}\label{Introduction}

Tools from information geometry (IG) have found a rich set of applications in the field of neuroscience. The book by Kass et al. \cite{kass2014Analysis} gives an excellent overview of the broad area. For approaches more specific to IG, see also work by Amari and Nagaoka \cite{amari2000Methods},
Amari \cite{amari2001Information}, 
Nakahara and Amari \cite{nakahara2002InformationGeometric}, 
Tatsuno et al. \cite{tatsuno2009InformationGeometric}, 
Nie and Tatsuno \cite{nie2012InformationGeometric}, 
Iwasaki et al. \cite{iwasaki2018Estimation}, 
Kass et al. \cite{kass2018Computational}, 
and recent work by Crosser and Brinkman \cite{crosser2024Applications}.
An especially important feature of the IG approach is the ability to model higher-order dependencies beyond simple pairwise associations. In this paper, we describe an investigation of these higher-order interactions in data from electroencephalography (EEG), a well-known, non-invasive technique that measures the electrical activity of the brain.

While our focus is on EEG data, the methodological foundation for this work comes established applications of IG to \emph{spike train} analysis. The human brain is a network of more than 80 billion cells -- called \emph{neurons} -- which interact with each other through electrochemical waves called \emph{action potentials}. The voltage of these electrochemical waves can be measured and recorded via sensitive electrodes placed near the active neurons. Partly due to the distinctive shape of these waves -- as measured on the voltage scale -- and because of their very short timescale, typically around a few milliseconds, action potentials are also called \emph{spikes}, and a time sequence of these neural spikes is called a \emph{spike train}. Since neurons display a characteristic `all or nothing' behaviour \cite{kandel2021Principles}, this means that there is almost no loss in representing the spike train as the cell being in two states: `on' $(1)$ for a very short period of time or `off' $(0)$. We can think of such data as being a realisation of a set of binary random variables $\{X_{i}\}$ indexed in some way, perhaps by time, experimental replication number, neuron label, or some combination of all of these. 

In Section \ref{Modelling Binary Random Vectors}, we explore the IG of models for such binary data. However, in this paper, we investigate an application of IG not to spike trains, but to data from EEG experiments. While spike trains reflect the activity of individual neurons as recorded by an electrode inserted into tissue, EEG records the aggregate electrical activity of large populations of neurons using electrodes placed on the scalp. EEG does not measure action potentials directly, but instead records the electric fields generated by the flow of currents from incoming connections to the neurons. When many neurons with similar orientations are synchronously active, their individual fields sum to produce a signal detectable at the scalp as a voltage fluctuation in the EEG trace \cite{schomer2011Niedermeyers}. EEG can be recorded using tens or hundreds of sensors, arranged across the surface of the scalp according to standardized systems, and is typically sampled at rates of 250--1000 Hz, providing high temporal resolution of neural dynamics. In EEG, the term \emph{electrode} refers to the sensor on the scalp which measures electrical activity, while the term \emph{channel} refers to the digital signal acquired by comparing the voltage at a given electrode with some reference point. The non-invasive nature of EEG and its relatively low cost have made it one of the most widely used methods for studying brain activity in humans. As the signal from EEG changes gradually and in an oscillatory manner, it does not lend itself to a binary representation as neatly as spike train data. If we wish to apply the IG models which have been successful in spike train analysis to EEG experiments, we must employ some rule that allows us to translate the signal to a 1 when certain conditions are met, and to a 0 otherwise. Different rules can be expected to capture different aspects of the signal. We discuss some of these methods of binarization in Section~\ref{Signal Binarization}.

In neuroscience, it is recognized that the brain is composed of many functionally specialized regions responsible for performing specific sets of computations \cite{kandel2021Principles}. Communication between these regions allows for the complex processing and integration of information necessary to produce behaviour and cognition \cite{bressler2010Largescale}. The organization of the brain can be characterized in terms of \emph{anatomical connectivity}, achieved by mapping the physical connections between regions \cite{sporns2005Human, elam2021Human}, or in terms of \emph{functional connectivity}. As defined by Friston \cite{friston1994Functional}, functional connectivity describes relationships between regions based on measures of their correlated activity without knowledge of the underlying anatomical structure. Observations of these statistical dependencies have resulted in the identification of many \emph{functional connectivity networks} that have been useful in explaining how the brain processes information. For an overview of these functional networks, see work by Bullmore and Sporns \cite{bullmore2009Complex}, Bressler and Menon \cite{bressler2010Largescale}, Yeo et al. \cite{thomasyeo2011Organization}, and Bassett and Sporns \cite{bassett2017Network}. For evidence of how these networks change as a result of aging, see the review by Damoiseaux \cite{damoiseaux2017Effects}. See work by Seeley et al. \cite{seeley2009Neurodegenerative} for examples of how these networks are affected in neurodegenerative disease. While these networks have been useful thus far, they are constructed from pairwise relations only. However, it is of scientific interest to understand if it is sufficient to focus only on pairwise dependencies, or if third or higher-order interactions play a role. See Battiston et al. \cite{battiston2020Networks, battiston2021Physics} for discussions on the importance of higher-order interactions in neuroscience and other fields. In this study, we applied IG methods to characterize third-order interactions in EEG data, investigating whether genuine higher-order dependencies beyond pairwise connectivity patterns could be extracted from real data.

\subsection{Modelling Binary Random Vectors}\label{Modelling Binary Random Vectors}

Joint distributions of $N$ binary random variables $X=(X_1,\dots,X_N)$ can be represented via the notation
\begin{equation*}
p_x := \Pr(X_1=x_1, \cdots, X_N=x_N) \in \Delta^{(K)} := \left\{ p_x \;|\; \forall x\in {\cal X}, \; p_x \ge 0, {\rm\; and\;} \sum_{x \in {\cal X}} p_x=1\right\},
\end{equation*}
with ${\cal X}$ being the set of $N$-dimensional binary vectors and $K := \left| {\cal X}\right| -1= 2^N-1$.  

The relative interior of the closed simplex $\Delta^{(K)}$ is a regular exponential family and the usual dual affine IG structure with its associated dual Pythagorean theorem can be applied. For example,  \cite{amari2001Information} uses this model in a spike train context. However, the faces of the simplex -- where a subset of probabilities $p_x$ can be zero -- will play an important role in the modelling. Hence, we extend the IG to the full closed simplex by viewing it as an extended exponential family: see Barndorff-Nielsen \cite{barndorff-nielsen2014Information}, Brown \cite{brown1986Fundamentals}, Csisz\'{a}r and Matus \cite{csiszar2005Closures}, and Anaya-Izquierdo et al. \cite{anaya-izquierdo2014When}. 

The reason for studying this closed space -- rather than an open manifold -- in the EEG or spike train context is that we would like to understand general dependence structures between the sets of binary responses. However, data from real experiments can be very sparse in terms of the observed counts of patterns of dependence. Given the relatively small number of replications typical of real-world experiments, it will often be the case that patterns of higher-order interactions are not observed frequently or at all. With such data, the purely empirical models will lie on lower dimensional faces of the simplex rather than in the interior. Furthermore, standard asymptotic tools such as having $\chi^2$ distributions for divergence measures may not be good approximations in this sparse context. 

The problem of inference given unobserved events has a long history, going back to Laplace's Rule of Succession and related methods based on adding small constants, which we employ in our analysis. These heuristics are similar to continuity corrections, such as Yates' correction \cite{yates1934Contingency}, which are commonly used to improve the quality of the normal approximation, particularly for the binomial case and in conditional and unconditional tests. An extensive literature addresses these and related approaches, see, for example, \cite{haber1982Continuity, brown2001Interval, emura2018Critical} and references therein.

\section{IG Measures}\label{IG Measures}

In this section, we describe the key IG relations necessary for our analysis. The general case and derivations of these relations are covered in detail by \cite{amari2001Information} and \cite{nakahara2002InformationGeometric}.

For the third-order case, we have three binary variables $X_1$, $X_2$, and $X_3$ whose set of joint probabilities $p(\boldsymbol{x})$, $\boldsymbol{x} = (x_1,x_2,x_3)$, $x_i = 0, 1$ is given by 8 probabilities. The joint probabilities of $\boldsymbol{p}$, given as a vector once an ordering has been selected, are
\begin{equation*}
	(p_{000} , p_{001} , p_{010} , p_{011} , p_{100} , p_{101} , p_{110}, p_{111}),
\end{equation*}
which are constrained to add to 1. Hence, the vector $\boldsymbol{p}$ lies in a 7-dimensional simplex, $\Delta^{(7)}$, whose interior is a manifold. One coordinate system of the interior of $\Delta^{(7)}$ is given by the expectation parameters $\boldsymbol{\eta}$,
\begin{equation*}
(\eta_{1},\eta_{2},\eta_{3},\eta_{12},\eta_{13},\eta_{23},\eta_{123})
\end{equation*}
\noindent{}where,
\begin{gather*}
\eta_{1} = E \{x_{1} = 1\} = p_{100}+p_{101}+p_{110}+p_{111} \, , \\[0.2em]
\eta_{2} = E \{x_{2} = 1\} = p_{010}+p_{011}+p_{110}+p_{111} \, , \\[0.2em]
\eta_{3} = E \{x_{3} = 1\} = p_{001}+p_{011}+p_{101}+p_{111} \, , \\[0.2em]
\eta_{12} = E \{x_{1}x_{2} = 1\} = p_{110}+p_{111} \, , \\[0.2em]
\eta_{13} = E \{x_{1}x_{3} = 1\} = p_{101}+p_{111} \, , \\[0.2em]
\eta_{23} = E \{x_{2}x_{3} = 1\} = p_{011}+p_{111} \, , \\[0.2em]
\eta_{123} = E \{x_{1}x_{2}x_{3} = 1\} = p_{111} \, . \\
\end{gather*}
This coordinate system then naturally extends to the boundary of the simplex. When referring to neural spiking, these $\eta$-coordinates represent the marginal and joint firing rates. In our case, the $\eta$-coordinates represent activity derived from EEG signals, so we say the $\eta$-coordinates represent the marginal and joint \emph{activation rates} of the binary signals.

In the interior of the simplex, the term $p(\boldsymbol{x})$ can be expanded on the log scale by
\begin{equation*}
	\log{p(\boldsymbol{x})} = \sum_{}\theta_{i}x_{i} + \sum_{}\theta_{ij}x_{i}x_{j} + \theta_{123}x_{1}x_{2}x_{3} - \psi ,
\end{equation*}
giving, in the interior, the natural parameters $\boldsymbol{\theta}$,
\begin{equation*}
	(\theta_{1},\theta_{2},\theta_{3},\theta_{12},\theta_{13},\theta_{23},\theta_{123}),
\end{equation*}
where,
\begin{gather*}
	\theta_{1} = \log{\frac{p_{100}}{p_{000}}} , \quad
	\theta_{2} = \log{\frac{p_{010}}{p_{000}}} , \quad
	\theta_{3} = \log{\frac{p_{001}}{p_{000}}} ,\\[1em]
	\theta_{12} = \log{\frac{p_{000}p_{110}}{p_{010}p_{100}}} , \quad
	\theta_{13} = \log{\frac{p_{000}p_{101}}{p_{001}p_{100}}} , \quad
	\theta_{23} = \log{\frac{p_{000}p_{011}}{p_{001}p_{010}}} ,\\[1em]
	\theta_{123} = \log{\frac{p_{001}p_{010}p_{100}p_{111}}{p_{110}p_{101}p_{011}p_{000}}} , \quad
	\psi = -\log{p_{000}},
\end{gather*}
which we call the $\theta$-coordinates. Each face of the simplex has a corresponding set of natural parameters determined by the support of the random vector. These are topologically connected via the polar dual of the simplex, see \cite{anaya-izquierdo2014When} for details.
We can also have mixtures of these $\eta$- and $\theta$-coordinates from two orthogonal foliations of the interior of $\Delta^{7}$:
\begin{gather*}
	\boldsymbol{\zeta}_{1} = (\eta_{1},\eta_{2},\eta_{3},\theta_{12},\theta_{13},\theta_{23},\theta_{123}), \\
	\boldsymbol{\zeta}_{2} = (\eta_{1},\eta_{2},\eta_{3},\eta_{12},\eta_{13},\eta_{23},\theta_{123}).
\end{gather*}
Using these mixed-cut coordinates, the Kullback-Leibler (KL) divergence between any two interior distributions $p(x)$ and $q(x)$ decomposes orthogonally as
\begin{gather*}
	D[p:q] = D[p:\bar{p}] + D[\bar{p}:\tilde{p}] + D[\tilde{p}:q],
\end{gather*}
where $p$, $\bar{p}$, $\tilde{p}$, and $q$ have mixed coordinates given by,
\begin{gather*}
	\boldsymbol{\zeta}_{2}^{p} = (\eta_{1}^{p},\eta_{2}^{p},\eta_{3}^{p},\eta_{12}^{p},\eta_{13}^{p},\eta_{23}^{p},\theta_{123}^{p}), \\[0.5em]
	\boldsymbol{\zeta}_{2}^{\bar{p}} = (\eta_{1}^{p},\eta_{2}^{p},\eta_{3}^{p},\eta_{12}^{p},\eta_{13}^{p},\eta_{23}^{p},\theta_{123}^{q}), \\[0.5em]
	\boldsymbol{\zeta}_{1}^{\tilde{p}} = (\eta_{1}^{p},\eta_{2}^{p},\eta_{3}^{p},\theta_{12}^{q},\theta_{13}^{q},\theta_{23}^{q},\theta_{123}^{q}), \\[0.5em]
	\boldsymbol{\zeta}_{1}^{q} = (\eta_{1}^{q},\eta_{2}^{q},\eta_{3}^{q},\theta_{12}^{q},\theta_{13}^{q},\theta_{23}^{q},\theta_{123}^{q}).
\end{gather*}
The superscripts $p$ and $q$ indicate whether that parameter comes from $p(x)$ or $q(x)$. When measuring how these two distributions differ, $D[p:\bar{p}]$ represents information from the triplewise interaction, $D[\bar{p}:\tilde{p}]$ represents information from the pairwise interaction, and $D[\tilde{p}:q]$ represents information from the modulation of the activation rates.

One important feature of this methodology is the flexibility it offers in selecting a null-hypothesis via the choice of $q(x)$. We can then test if the estimated third-order interaction differs significantly from the null using the test statistic
\begin{gather*}
	\lambda = 2ND[p:\bar{p}] \sim \chi^2(1),
\end{gather*}
given a sufficiently large $N$.

We can relate these IG measures to analyze the relation between random vectors $X$ and $Y$ via the mutual information
\begin{gather*}
	I(X,Y) = E_{p(Y)}[ \, D \, [ \, p(X|Y):p(X) \, ]]
\end{gather*}
which can be decomposed in a similar manner to the KL divergence, yielding
\begin{gather*}
	I(X,Y) = I_1(X,Y) + I_2(X,Y) + I_3(X,Y), \\[1em]
	I_1(X,Y) = E_{p(Y)}[ \, D \, [ \, \boldsymbol{\zeta}_{1}^{\tilde{p}}(X,y) : \boldsymbol{\zeta}_{1}^{q}(X) \, ]], \\[0.5em]
	I_2(X,Y) = E_{p(Y)}[ \, D \, [ \, \boldsymbol{\zeta}_{2}^{\bar{p}}(X,y) : \boldsymbol{\zeta}_{1}^{\tilde{p}}(X,y) \, ]], \\[0.5em]
	I_3(X,Y) = E_{p(Y)}[ \, D \, [ \, \boldsymbol{\zeta}_{2}^{p}(X|y) : \boldsymbol{\zeta}_{2}^{\bar{p}}(X,y) \, ]].
\end{gather*}
$I_1$ is the mutual information from the modulation of the activation rates, $I_2$ from the pairwise interaction, and $I_3$ from the triplewise interaction. As noted in Section~\ref{Modelling Binary Random Vectors}, real data can be sparse and result in probability estimates on the boundary of the simplex, requiring care when applying this IG methodology. For the real data analysed here, the Laplace adjustment of adding a small positive term to probability estimates to keep them away from the boundary was effective for the analysis. Sparser problems will likely require more sophisticated IG tools for addressing boundary cases.

\section{IG Analysis}\label{IG Analysis}

\subsection{EEG Dataset}\label{Motor Imagery Dataset}

A publicly available EEG dataset was used in this analysis, accessed through OpenNeuro  \cite{triana-guzman2024EEG}. This dataset was generated from human EEG recordings taken during a motor imagery (imagined movement) experiment performed by Triana-Guzman et al. \cite{triana-guzman2022Decoding}. The goal of this experiment was to identify patterns of EEG activity that indicated whether a participant was imagining standing up from a sitting position, or sitting down from a standing position. To isolate the EEG activity associated with motor imagery, the researchers compared trials involving the motor imagery task with those involving a control condition. Successfully identifying these patterns of activity could prove helpful to those with lower limb disabilities. A potential application would be integrating a brain-computer interface (BCI), which detects EEG activity indicative of an intended postural change, with an assistive device that could aid a person in sitting down or standing up.

\begin{figure}[ht]
	\centering
	\includegraphics{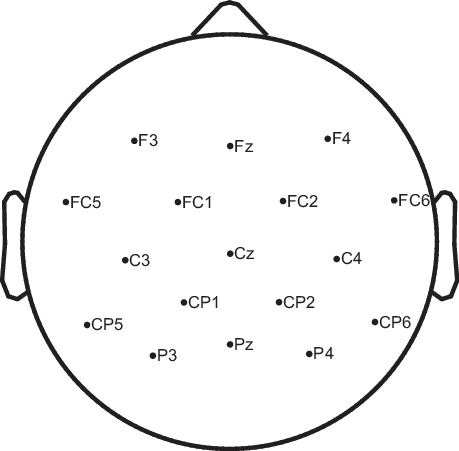}
	\caption{Positions of electrodes on the scalp with 10-20 labelling. The letters `F', `C', and `P' stand for `frontal', `central', and `parietal', respectively. Electrode labels ending in `z' indicate those placed along the midline}
	\label{fig1}
\end{figure}

These EEG data were recorded from 32 healthy participants (16 female) using 17 electrodes placed over frontal, central, and parietal regions according to the International 10-20 System \cite{sharbrough1991American} and sampled at a rate of 250 Hz (Fig.~\ref{fig1}). The data were minimally processed using a band-pass filter with cutoff frequencies of 0.01 and 60 Hz \cite{triana-guzman2022Decoding}.

The experiment was separated into offline and online phases. The offline phase consisted of six blocks, with breaks in between to avoid fatiguing the participants. In three blocks, the participant was sitting (condition A); in the other three blocks, they were standing (condition B). The order of the blocks was chosen by the participant. In each block, the participant was guided through 30 trials by a graphical user interface (GUI) displayed on a screen in front of them: 15 trials of the motor-imagery task and 15 trials of the associated idle-state task, presented in a pseudorandom order. EEG was recorded continuously through all blocks. As detailed by \cite{triana-guzman2022Decoding}, each trial consisted of the following four steps:
\begin{enumerate}
	\item \textbf{Fixation (4 s)} Participants were asked to stare at the fixation cross presented on-screen while avoiding blinking and other movements.
	\item \textbf{Observation (3 s)} The GUI displayed a figure showing which task the participant will need to carry out in the subsequent step. Participants were to refrain from beginning the task until step 3.
	\item \textbf{Imagination (4 s)} The participant carried out the mental task indicated in step 2. The tasks consisted of the following:
	\begin{itemize}
		\item \textbf{Motor Imagery A (motorA)} While sitting, imagine carrying out a sit-to-stand movement.
		\item \textbf{Idle State A (idleA)} While sitting, do not imagine carrying out a sit-to-stand movement.
		\item \textbf{Motor Imagery B (motorB)} While standing, imagine carrying out a stand-to-sit movement.
		\item \textbf{Idle State B (idleB)} While standing, do not imagine carrying out a stand-to-sit movement.
	\end{itemize}
	\item \textbf{Rest (4 s)} The word `Descanso' was displayed on screen (Spanish for `rest'). The participant was allowed to blink and move the head freely during this period.
\end{enumerate}

Over the six blocks, a total of 45 trials of each task (motorA, idleA, motorB, idleB) were completed during the offline phase. Following the offline phase, an online phase of the experiment was administered where the researchers used a machine learning model to classify the imagined movement in near real time. This phase consisted of two additional blocks that followed the same design as in the offline phase but with a modified step 3 (the imagination task). In the online phase, the imagination step was of variable length (from 3 s to 15 s) and the participant received continuous visual feedback from the output of the machine learning model. Due to variable trial length and altered input to the participant, we restricted our IG analysis to data collected in the offline phase only. See Fig.~\ref{fig16} in Appendix A for an overview of the experimental design.

The original experimental protocol was approved by the ethics committee of the Universidad Antonio Nari\~{n}o \cite{triana-guzman2022Decoding}. Second-hand use of the dataset was approved by the University of Alberta Research Ethics Board (Pro00121346).

\subsection{Data Preprocessing}\label{Data Preprocessing}
Preprocessing of the motor imagery data was done to remove artifacts, filter the data into different frequency bands, and to extract relevant data epochs. All preprocessing was completed using the EEGLAB toolbox for MATLAB \cite{delorme2004EEGLAB} (version 2025.0.0). This dataset included electrode labels following the 10-20 system but did not include precise electrode locations. Electrode locations for the given labels were imported from the standard 10-5 template in EEGLAB. See Appendix A for more details on initial filtering and artifact removal.

\subsubsection{Band-Pass Filtering}\label{Band-Pass Filtering}
Conventional time-domain analyses of EEG data examine signals within recognized frequency bands, where differential activity in those bands is associated with various types of cortical processing or conscious states. Broadly speaking, as cognitive activity increases so does the dominant EEG frequency. During sleep, for example, low-frequency rhythms tend to dominate. Table~\ref{table1} lists the typical frequency bands of interest along with their general associations to behaviour and neural processing. This is not an exhaustive list however, and patterns of activity within these frequency bands are complex and non-uniform across the brain \cite{schomer2011Niedermeyers}. We filtered the EEG signals into different frequency bands using band-pass FIR filters with the cutoff frequencies listed in Table~\ref{table1}. Beta was split into low, middle, and high sub-bands consistent with the original analysis by \cite{triana-guzman2022Decoding}.

\begin{table}[h]
	\centering
	\caption{Frequency bands of interest and some associated phenomena}
	\label{table1}
	\begin{tabular}{c  c  c}
		\toprule
		\quad Frequency Band \quad & \quad Range (Hz) \quad & Associated Phenomena \\
		\midrule
		Delta & 0.5--4 & Deep sleep, anesthesia \\
		\midrule
		Theta & 4--8 & Rapid eye movement (REM) sleep, \\
		&      & memory consolidation and retrieval \\
		\midrule
		Alpha & 8--12 & Relaxed wakefulness, attentional control \\
		\midrule
		\multirow{3}{*}{Beta} 
		& low 12--16 & Focused attention, \\
		& mid 16--20 & motor programming \\
		& high 20--30 & \\
		\midrule
		Gamma & 30--60 & Sensory processing, perception \\
		\bottomrule
	\end{tabular}
\end{table}

\subsubsection{Extracting Epochs}\label{Extracting Epochs}
\begin{figure}[ht]
	\centering
	\includegraphics{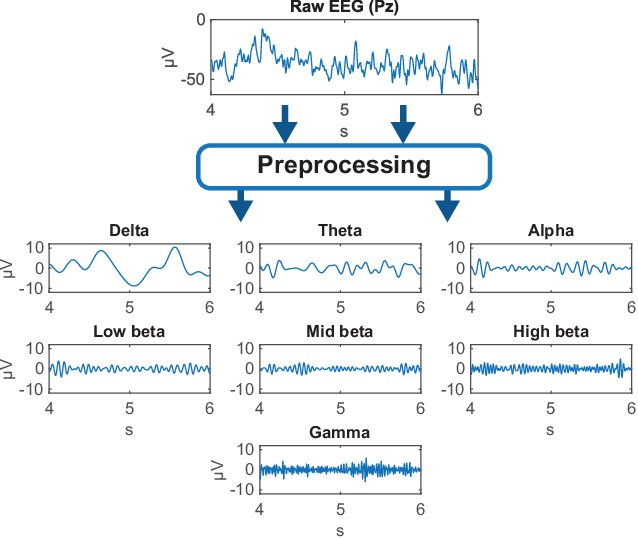}
	\caption{Outputs of preprocessing showing a 2-second sample from channel Pz}
	\label{fig2}
\end{figure}

The relevant portion of each trial was extracted using the event markers supplied with the dataset. Epochs extended from 7 s before the onset of the imagination task (step 3) to 4 s after onset. This 11-second period contains the fixation, observation, and imagination phases while excluding the noisy resting phase. Epochs were sorted based on the condition indicated in the trial: motorA, idleA, motorB, or idleB. This provided 45 11-second trials per task category for each of the 32 participants. Figure~\ref{fig2} summarizes the outputs from the preprocessing pipeline.

\subsection{Signal Binarization}\label{Signal Binarization}
As discussed in Section~\ref{Introduction}, it was necessary to first reduce the EEG signals to a binary representation before proceeding with our analysis. In theory, this discretization could be done using any number of states \cite{amari2001Information}. However, as the number of discrete states increases, so too does the amount of data required to accurately estimate the likelihood that those states occur. To avoid this curse of dimensionality, we limited our discretization to only two states.

Even in this simple case, the discretization of a continuous oscillatory signal can be accomplished in a variety of ways that capture different aspects of the signal, affecting downstream analysis. We used three straightforward, complementary binarization methods: two sign-based methods and a threshold-based method. We will refer to these as the Sign, Diff, and Power methods throughout this manuscript.

\subsubsection{Sign Binarization}\label{Sign Binarization}
In this method, EEG signals were binarized directly by sign. Values $>0$ were coded as 1, else coded as 0. This method captured coarse phase information while ignoring amplitude.

\subsubsection{Diff Binarization}\label{Diff Binarization}
Here, the first difference of the signal was computed, then binarized by sign ($>0$ coded as 1). Binary signals were appended with a 0 to preserve signal length. This captured coarse phase transition patterns.

\subsubsection{Power Binarization}\label{Power Binarization}
Signal amplitudes were squared to obtain power, the upper envelope was estimated using a 30-sample moving average, and the \textit{z}-scores of the envelope were computed. \textit{Z}-scores $>1$ were coded as 1, otherwise 0. The \textit{z}-scores for each EEG channel were calculated using that channel's envelope values across all trials within both conditions of the seated or standing blocks (motorA and idleA; motorB and idleB), allowing for direct comparison between motor imagery and idle-state conditions. This method captured periods of high amplitude while being agnostic to phase.

\begin{figure}[ht]
	\centering
	\includegraphics{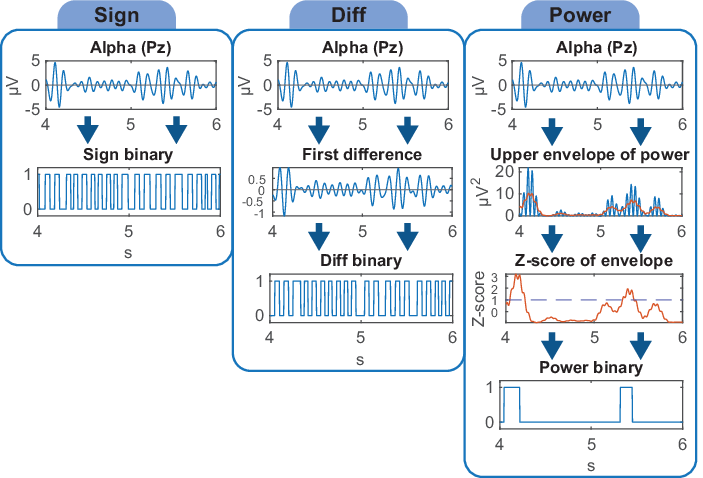}
	\caption{Summary of Sign, Diff, and Power methods using a 2-second sample in the alpha frequency band from channel Pz}
	\label{fig3}
\end{figure}

Figure~\ref{fig3} summarizes the operations of each binarization method. While the Sign and Diff methods produce similar binary signals, differing primarily by a phase shift, their outputs increasingly diverge in lower frequency bands.

\subsection{Computing IG Measures}\label{Computing IG Measures}
Using the binarized signals, we computed third-order IG measures (defined in Section~\ref{IG Measures}) for all unique combinations of three EEG channels. The probability vector for each triplet was calculated based on counts of discrete events within 1-second, non-overlapping windows (8 possible binary state combinations per triplet). The probability vectors from each window were then averaged across all trials within separate conditions. Laplace smoothing was performed at this step to guarantee non-zero probabilities for cases where a discrete event was never observed,
\begin{gather*}
	\hat{\theta_{i}} = \frac{x_{i}+\alpha}{N + \alpha d} \:, \quad (i=1,\ldots,d), \quad \alpha=1,
\end{gather*}
where $x_{i}$ is the count of observations for a specific event, $N$ is the total number of observations, and
\begin{gather*}
	\hat{\theta_{1}} = p_{000} \: , \: \hat{\theta_{2}} = p_{001} \: , \ldots , \: \hat{\theta_{8}} = p_{111} \: .
\end{gather*}
The KL divergence was computed between distributions from the motor imagery and corresponding idle-state conditions for each 1-second window over the duration of the trial. These divergences were then decomposed into first-order, second-order, and third-order components, and the mutual information was determined as described in Section~\ref{IG Measures}. While Laplace smoothing might be considered a rather ad hoc approach for dealing with sparse probabilities, it was found to be both simple and effective in this particular problem. At higher levels of sparsity, for example when looking for interactions beyond third-order, we expect that the full geometric structure of the extended multinomial models will play an important role.

\section{Surrogate Data Testing}\label{Surrogate Data Testing}
\subsection{Surrogate Datasets}\label{Surrogate Datasets}

\begin{figure}[ht]
	\centering
	\includegraphics{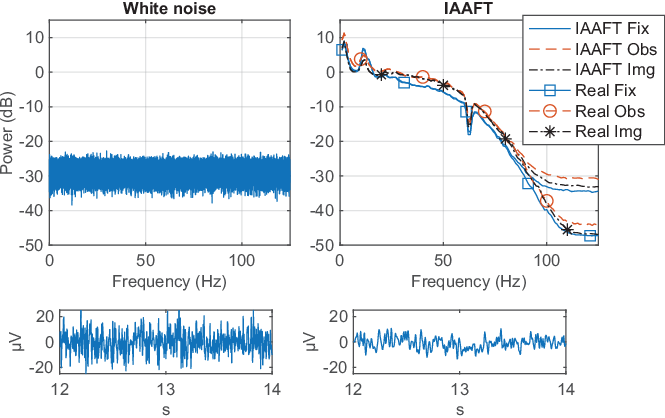}
	\caption{Welch power spectrum estimates for white noise and IAAFT surrogate data. Power spectrums represent signals from one EEG channel and one participant. For the IAAFT surrogate data, power spectrums are shown for separate trial phases. ``Fix", ``Obs", and ``Img" refer to fixation, observation, and imagination phases, respectively. Spectrums were the average over all motorA and motorB trials (90 total trials). During the fixation phase, higher power was seen in the alpha frequency range, and lower power in mid beta and higher bands as compared to observation and imagination phases. Spectrums of real data for the corresponding participant are plotted for comparison. The power spectrums of real and IAAFT data were nearly identical until approximately 80 Hz where they began to diverge, which was beyond the upper limit of our analysis (gamma, 30--60 Hz). Sample signals are shown at bottom for both types of surrogate data}
	\label{fig4}
\end{figure}

To investigate the level of spurious information produced by our analysis and to support our definition of the null hypothesis, surrogate datasets were created by replacing EEG signals with white noise or phase-randomized signals. All structural aspects of the motor imagery dataset were maintained including number of channels, number and length of trials, number of participants, and overall data organization. The surrogate data began and ended with 20 s periods of null activity to reflect EEG recorded before and after the experiment. Sitting and standing trial blocks were then simulated in a pseudorandom order with a 20 s period between blocks to simulate breaks. Each block consisted of 30 trials, emulating the configuration of motor imagery and idle-state tasks. Each trial was a total of 15 s long, encompassing the equivalent of fixation, observation, imagination, and rest periods. As with the real data, the rest period of each trial was omitted during the selection of epochs. In the first surrogate dataset, white noise was used throughout to investigate the effects that may emerge from purely random structure. These data were generated 32 times with different random seeds to represent the 32 participants in the original dataset. In the second surrogate dataset, an iterated amplitude adjusted Fourier transform (IAAFT) \cite{schreiber1996Improved} procedure was used on signals from individual participants to construct surrogate data with the same power spectrum and distribution of amplitudes, but with randomized phases. Within the data for each participant, IAAFT was used on each trial phase separately to achieve noised signals that preserved changes in the power spectrum seen in different trial phases.  Figure~\ref{fig4} shows power spectrum estimates and sample signals from the two types of surrogate data. These data were band-pass filtered into the same frequency bands as with the original dataset before proceeding with the IG analysis.

\subsection{Surrogate Data Results}\label{Surrogate Data Results}
We first examined the decomposed mutual information of each surrogate dataset by pooling results across our 32 simulated participants. Overall, the IAAFT data produced higher information values than white noise, and the Power method yielded the highest values among the three binarization methods.

\begin{figure}[ht]
	\centering
	\includegraphics{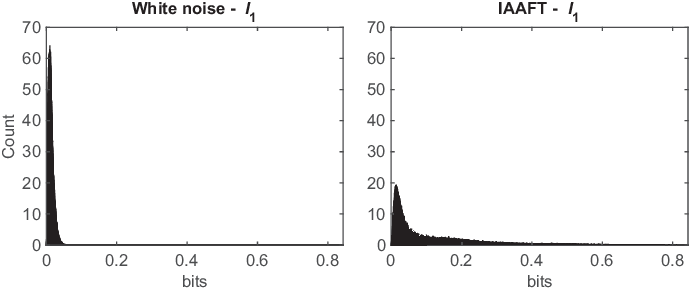}
	\caption{Distributions of $I_1$ in the alpha frequency band from the white noise and IAAFT datasets using the Power method, pooled across all iterations. Maximum values of $I_1$ were 0.06617 bits for white noise, 0.8435 bits for IAAFT. Much higher values of $I_1$ were found in the IAAFT case because first-order power dynamics were preserved in the signals}
	\label{fig5}
\end{figure}

The maximum mutual information values occurred for $I_1$ using the Power method on IAAFT data (Fig.~\ref{fig5}). Importantly, these elevated $I_1$ values were not spurious but rather expected given the IAAFT procedure. The Power method captures periods of high signal amplitude, and $I_1$ reflects changes in activation rates -- the frequency of these high-amplitude periods. Because IAAFT preserved the power spectrum differences between trial phases and conditions, it necessarily preserved the distribution of high-amplitude periods, resulting in higher first-order information that accurately reflects these preserved spectral properties.

\begin{figure}[ht]
	\centering
	\includegraphics{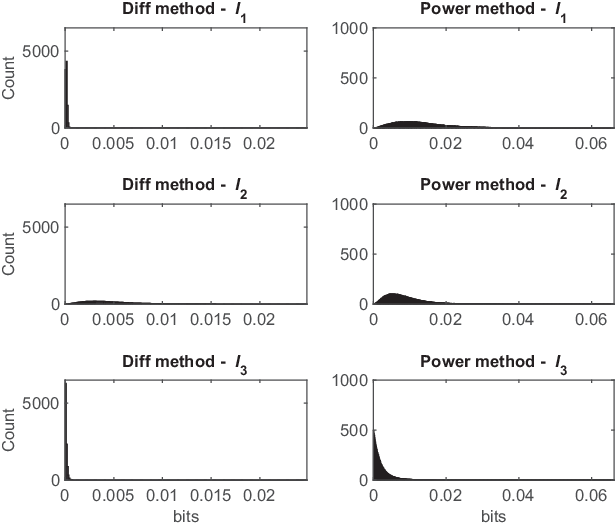}
	\caption{Distributions of $I_1$, $I_2$, and $I_3$ in alpha using the Diff and Power methods with white noise, pooled across all iterations. The Diff method showed a tendency for higher information in the second order, with $I_2$ extending up to 0.02483 bits, while maxima for $I_1$ and $I_3$ were 0.0007507 bits and 0.001476 bits, respectively. The Power method showed a decreasing trend as the order of interaction increased. Note the difference in x and y axes limits between the Diff and Power methods}
	\label{fig6}
\end{figure}

With the Power method, $I_1 > I_2$ and $I_2 > I_3$ in all frequency bands using both white noise and IAAFT data. This was not the case for the sign-based binarization methods. Both the Sign and Diff methods produced much smaller values of $I_1$ and $I_3$ than $I_2$. This effect is less pronounced in the delta frequency range, but more obvious in higher frequency bands. We can see an example of this in the alpha band using the Diff method in Fig.~\ref{fig6}. 

\begin{figure}[ht]
	\centering
	\includegraphics{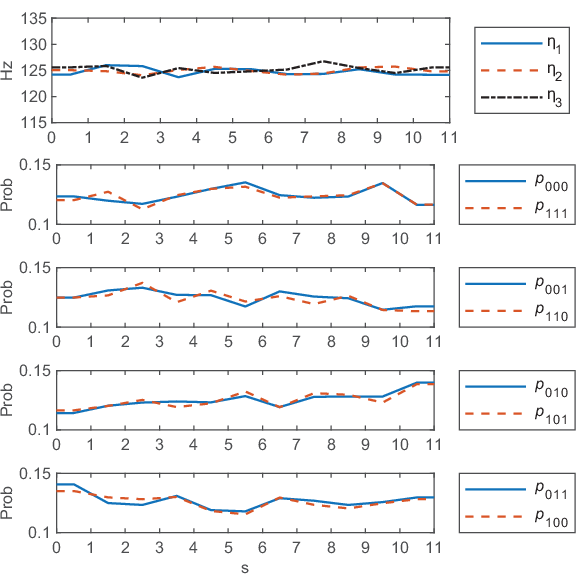}
	\caption{Top panel shows the marginal activation rates over the course of the 11-second trials. The activation rates deviate only slightly from 125 Hz (half the sampling rate). Lower panels show how opposing probabilities are highly correlated over the course of the trials. All data shown are from one simulated participant using white noise filtered for alpha. Values are the mean of the 45 motorA trials}
	\label{fig7}
\end{figure}

This biased capture of second-order interactions highlights a shortcoming of applying sign-based binarization methods to oscillatory signals. As $I_1$ depends on the modulation of the activation rates $\eta_i$, this information will be low because $\eta_i$ tends not to deviate from a value of half the sampling rate of an oscillatory signal (Fig.\ref{fig7} top panel). Values of $I_3$ are low because probabilities of opposite binary state combinations (e.g. $p_{001}$ and $p_{110}$) tend to be highly correlated, shown in the lower panels of Fig.~\ref{fig7}. In the calculation of $\theta_{123}$, these opposing probabilities appear in the numerator and denominator of the likelihood ratio, forcing the value of $\theta_{123}$ toward zero.

\section{Motor Imagery Data Results}\label{Motor Imagery Data Results}
\begin{figure}[ht]
	\centering
	\includegraphics{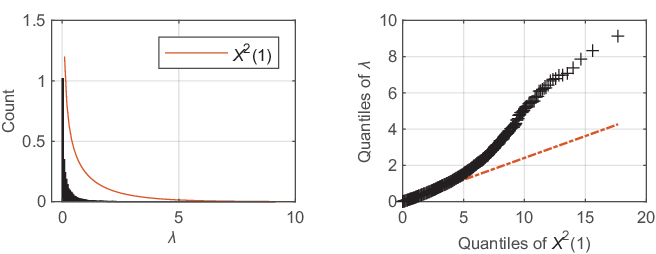}
	\caption{The left panel shows the distribution of $\lambda$ from the motorA/idleA conditions for one participant compared to $\chi^2$(1). The right panel shows the quantiles of $\lambda$ plotted against the quantiles of $\chi^2$(1), confirming a poor fit}
	\label{fig8}
\end{figure}

Given the limitation of the sign-based binarization methods in capturing third-order interactions discussed above, we focused on the results provided by the Power method when applying our analysis to the motor imagery dataset. To determine which channel triplets provide statistically significant third-order mutual information about the task condition (motor imagery or idle-state), we can use the test statistic $\lambda$ to evaluate each of the contributing KL divergences. However, perhaps as expected with only 45 trials, we found that the $\chi^2$ approximation was not a good fit to the null distribution, shown in Fig.~\ref{fig8}. The potential failure of standard asymptotic null distributions in this sparse context is one of the reasons we focused on the null behaviour with simulated data (Section~\ref{Surrogate Data Testing}).

\begin{figure}[ht]
	\centering
	\includegraphics{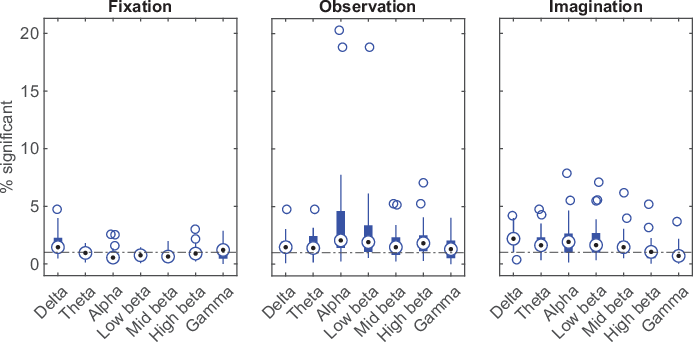}
	\caption{The percentage of triplets with significant $I_3$ ($p<0.01$) was computed by frequency band for each participant in different trial phases. These box plots show the median and interquartile range (IQR) of these percentages across participants in each trial phase. The threshold for outliers was $1.5*$IQR. Triplets with significant $I_3$ during the fixation phase can be considered false positives, while those in the observation and imagination phases would be considered true positives. The horizontal line across all plots corresponds to 1\%. Median values are closer to 1\% and have smaller IQRs in the fixation period compared to the observation and imagination periods}
	\label{fig9}
\end{figure}

To test for significance, we compared mutual information values from the motor imagery data to the null distributions produced by our IAAFT surrogate data. For third-order information, these null distributions were constructed for separate participants, frequency bands, and trial phases. We calculated $p$-values using these nulls and rejected any channel triplets with $p > 0.01$. During the fixation period of each trial, the participant did not yet know whether they would be instructed to complete a motor imagery or idle-state task. As such, the distribution of mutual information values in the fixation period should closely match those of the null distribution. Given our choice of threshold, the number of significant triplets found during the fixation period (false positives) should be approximately 1\% if the null distribution is appropriate. Figure~\ref{fig9} shows a comparison of the percentage of significant channel triplets identified in each trial phase across participants.

\begin{figure}[ht]
	\centering
	\includegraphics{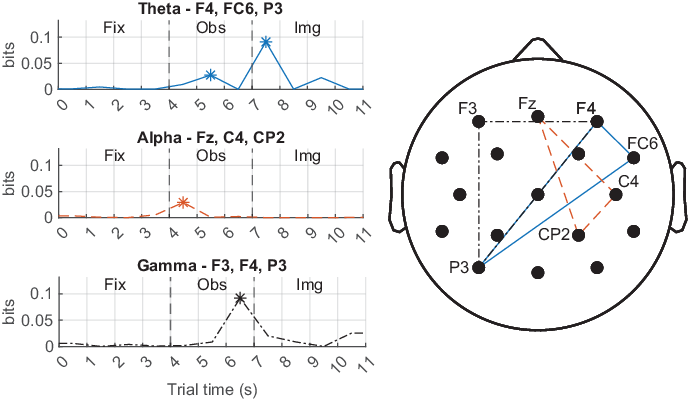}
	\caption{Left panels show $I_3$ over time for three channel triplets from one participant. One triplet is shown from theta, alpha, and gamma. The labels of corresponding channels involved in the triplets are also listed. ``Fix", ``Obs", and ``Img" refer to fixation, observation, and imagination phases of the trial. Points where the mutual information was significant are indicated by $*$ ($p<0.01$). There was low information during the fixation period and brief peaks during observation and imagination phases where significant information was found. The temporal dynamics of third-order information differed between frequency bands and triplets. Right panel shows the spatial organization of the plotted triplets on the scalp. The triplets in theta and gamma share an edge from F4 to P3}
	\label{fig10}
\end{figure}

An advantage of performing our IG analysis within 1-second windows, as described in Section~\ref{Computing IG Measures}, is that we can measure how the mutual information changes over time. Observing the temporal dynamics of information may provide insight into the role of higher-order interactions. Figure~\ref{fig10} shows the change in mutual information over time for three examples of channel triplets along with the spatial organization of those triplets.

\section{EEG Forward Model and Information Attenuation}\label{EEG Forward Model and Information Attenuation}
Third-order mutual information identified in the motor imagery data, while significant, was small. This was to be expected due to both the sparsity inherent in this type of analysis and the fact that any interaction between brain regions would not be perfectly captured by analyzing EEG recorded at the scalp. To obtain some measure of the reduction in information inherent in our analysis and thereby gain insight into the importance of the statistically significant results, we performed a validation exercise using a forward model. A forward model in EEG describes how the electric fields generated by current sources in the brain will be measured on the surface of the scalp. This is done by modelling the location of a source in the brain, the location of a sensor on the scalp, and the conductivity of the biological media that lie between them. Forward models are typically used in conjunction with inverse methods to estimate source activity from scalp recordings. In our approach, we reversed this process by beginning with synthetic source signals, where we assigned known second- and third-order dependencies, then computed the EEG signals expected at the electrodes. We then evaluated whether our IG analysis could reliably identify a known dependency at the source level based on the scalp signals, with varying levels of background noise. This validation approach was important in determining to what degree information is attenuated due to volume conduction effects and the sensitivity of the analysis to interaction strength relative to background activity.

This validation exercise was largely successful as we confirmed that our analysis could detect dependencies between sources in the brain based on the corresponding EEG signals. We also quantified the information loss at different steps in the analysis which provided insight as to where the greatest improvements could be achieved through optimization. The remainder of this section discusses the methodology of the forward model, implementation of the known dependencies, and detailed results.

Forward modelling was completed using the FieldTrip toolbox \cite{oostenveld2011FieldTrip} (version 20250523) for MATLAB. The forward model was produced using the boundary element method (BEM) \cite{oostendorp1991Potential} with the standard BEM head model \cite{oostenveld2003Brain}. This was a 3-shell model (brain, skull, scalp) based on the Colin 27 average brain \cite{holmes1998Enhancement}. Sources were positioned following a regular grid pattern with 10 mm spacing. This resulted in 1608 sources in the brain. The electrode positions used were the same as the 17 electrodes in the motor imagery dataset. This synthetic validation dataset followed the same structure as the motor imagery dataset so that comparisons could be drawn, and was produced using the same procedure as the surrogate datasets, except that only seated trial blocks and only one participant were simulated (motorA and idleA trials). Throughout this section, we look only at the results using the Power binarization method.

\begin{figure}[ht]
	\centering
	\includegraphics{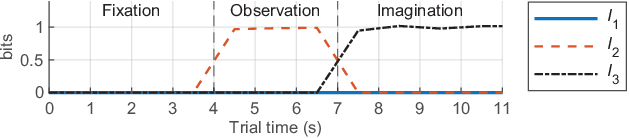}
	\caption{The change in $I_1$, $I_2$, and $I_3$ components over the course of simulated trials using dependency rules applied directly to binary signals. There was no information during the fixation phase when all signals behaved randomly, near perfect capture of second-order information during the observation phase when signals 2 and 3 were correlated, and also of third-order information during the imagination phase when the three signals had a deterministic relationship}
	\label{fig11}
\end{figure}

Before simulating EEG with a forward model, we first verified that we could enforce known second- and third-order dependencies that were identified as expected using our analysis. We began by applying dependency rules to binary signals directly. Second-order dependence was produced by simply having two target signals be perfectly correlated. Third-order dependence consisted of the elementary case where a third control signal determined whether the two target signals were correlated or anticorrelated. During the simulated idleA trials, all three binary signals had random activity for the duration of the trial. During the simulated motorA trials, interactions between the three signals changed depending on the phase of the trial. During the fixation phase, all three signals had random activity. During the observation phase, signals 2 and 3 (target signals) were correlated irrespective of the state of the control signal (second-order case). During the imagination phase, if the control signal was high (1), signals 2 and 3 were correlated. If the control signal was low (0), signals 2 and 3 were anticorrelated (third-order case). Figure~\ref{fig11} shows how information was captured over the duration of the trial.

\begin{figure}[ht]
	\centering
	\includegraphics{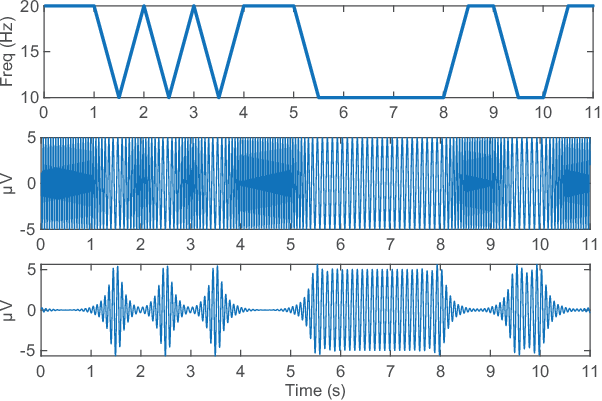}
	\caption{The top panel shows the frequency modulation vector of signal 1 over the course of a single trial. This signal oscillated randomly between 10 Hz and 20 Hz at 0.5 s intervals. The middle panel shows the sinusoidal signal generated from the frequency modulation vector with periods of 10 Hz and 20 Hz activity. The bottom panel shows the signal after band-pass filtering for alpha (8--12 Hz), leaving only the 10 Hz portion of the signal}
	\label{fig12}
\end{figure}

To implement these dependency rules in oscillatory source signals, we manipulated an artificial amplitude coupling interaction. To generate source signals, we started with frequency modulation vectors that described how the frequency switched between two discrete states: high (20 Hz) and low (10 Hz). This high or low state was assigned at 0.5 s intervals and points within those intervals were linearly interpolated to produce smooth frequency transitions. These frequency modulation vectors were then used to generate sinusoidal source signals, with random phases, whose frequency varied between 10 Hz and 20  Hz. These signals were then band-pass filtered for alpha (8--12 Hz). This meant that a transition of the frequency modulation vector to 10 Hz resulted in an increase in power in alpha, while transitions to 20 Hz resulted in a decrease in power in alpha (Fig.~\ref{fig12}).

\begin{figure}[ht]
	\centering
	\includegraphics{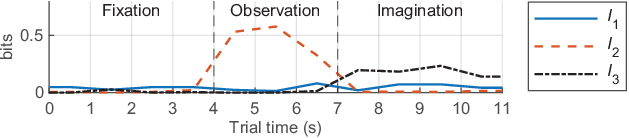}
	\caption{The change in $I_1$, $I_2$, and $I_3$ components over the course of simulated trials using oscillatory signals with enforced amplitude coupling relationships. There was no information during the fixation phase when all signals behaved randomly. There was much reduced information during the observation and imagination phases as compared to applying these dependency rules directly to binary signals. This effect was more pronounced for $I_3$ than $I_2$}
	\label{fig13}
\end{figure}

The frequency switching patterns were controlled by the dependency rules described above. During the idleA trials, all signals randomly switched between 10 Hz and 20 Hz for the duration of the trial. During the fixation phase of the motorA trials, switching of all signals was again random. During the observation phase, the control signal switched randomly while the target signals switched in a correlated manner (second-order case). During the imagination phase, if the control signal was switched to 10 Hz, the target signals were correlated; if the control signal was switched to 20 Hz, the target signals were anticorrelated (third-order case). These signals were binarized using the Power method and the IG analysis was then performed at this source level to determine how well the known interactions were identified when translated to oscillatory signals. Figure~\ref{fig13} shows the reduction in information capture associated with this translation.

\begin{figure}[ht]
	\centering
	\includegraphics{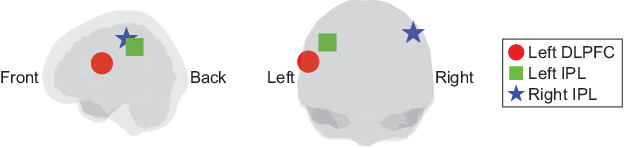}
	\caption{Locations in the brain used for the three source signals. Signal 1 was assigned to the left dorsolateral prefrontal cortex (DLPFC). Signals 2 and 3 were assigned to left and right inferior parietal lobules (IPL), respectively}
	\label{fig14}
\end{figure}

To produce simulated EEG from these three source signals, we first needed to assign these sources positions within the brain. Biologically plausible source locations were determined from the results of a functional magnetic resonance imaging (fMRI) study by Guillot et al. \cite{guillot2009Brain} that looked at the differential activation of brain regions during a finger-touching motor imagery task. Based on this study, we identified the three cortical areas that had the greatest activation (highest $t$-value) in the kinesthetic imagery versus perceptual control (rest) condition. These regions corresponded to the left and right inferior parietal lobules and the left dorsolateral prefrontal cortex. The unfiltered source signals were assigned by choosing the source position in our forward model that was nearest the coordinates of peak activation provided in \cite{guillot2009Brain}. These coordinates could not be mapped one-to-one to our head model as it is based on the Colin 27 anatomical data while \cite{guillot2009Brain} reports coordinates in the MNI152 space. However, these coordinates provide sufficient guidance for assigning biologically plausible source locations. The control signal was assigned to the source in the dorsolateral prefrontal cortex, while the target signals were assigned to sources in the left and right inferior parietal lobules (Fig.~\ref{fig14}). The dipoles associated with these sources were oriented orthogonal to the cortical surface. This is consistent with the orientation of pyramidal neurons in the cortex, whose activity makes up the majority of recorded scalp EEG signals \cite{schomer2011Niedermeyers}.

\begin{figure}[ht]
	\centering
	\includegraphics{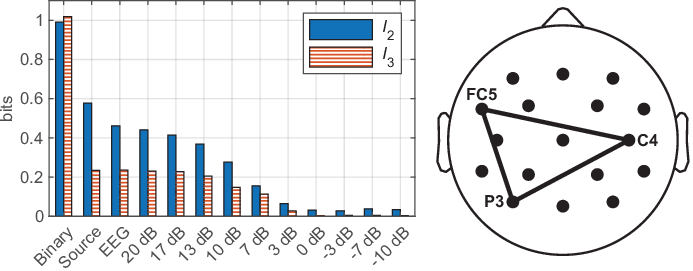}
	\caption{The left panel shows a comparison of the maximum $I_2$ and $I_3$ when applying dependency rules directly to binary signals (binary), translating those rules to oscillating signals (source), projecting those source signals to scalp electrodes with no noise (EEG), and projecting those signals to scalp electrodes with varying SNRs (values in dB). The right panel shows the spatial organization of EEG channels where the maximum values of $I_2$ and $I_3$ were found. These channels corresponded to those closest to the positions of the source signals. All comparisons refer to information values from this channel triplet}
	\label{fig15}
\end{figure}

After assigning our source signals to their positions in the brain, we calculated the EEG signals of our 17 channels with varying levels of noise. In the first case, the only signals present were the three assigned sources; all non-signal sources in the brain were silent. In subsequent cases, all non-signal sources produced pink noise of appropriate amplitude to achieve the desired signal-to-noise ratio (SNR). In these noised cases, non-signal sources were assigned random dipole orientations and SNRs varied from $-10$ dB to $20$ dB. We then band-pass filtered the EEG signals for alpha and carried out our IG analysis. SNRs were calculated based on power in the alpha band, and pink noise was chosen to mimic the $1/f$ spectral properties of EEG.  Figure~\ref{fig15} shows how information capture is reduced with decreasing SNR. The largest decrease in information occurred when moving from binary signals to oscillating signals, with $I_3$ being more affected than $I_2$. However, maximum $I_3$ was then relatively constant when projecting those source signals to the scalp electrodes and with low levels of pink noise. $I_2$ gradually decrease with the change from source-level signals to EEG and with increasing levels of pink noise. With higher levels of noise, both maximum $I_2$ and maximum $I_3$ were greatly diminished.

\section{Discussion}\label{Discussion}
This work demonstrates that IG methods, previously applied to discrete spike train data, can successfully detect third-order interactions in continuous EEG signals through appropriate binarization approaches.

Forward modelling revealed that the greatest reduction in information capture occurred when translating enforced dependencies from binary signals to oscillatory signals, rather than during projection from brain sources to scalp electrodes. This suggests that optimized binarization methods could substantially improve information capture, though the relative merits of different binarization techniques for continuous neural signals remains an open methodological question. Our conservative surrogate-based significance testing approach, using phase-randomized data with preserved spectral properties, successfully controlled false positive rates as evidenced by near-chance detection levels during fixation periods.

The forward model represented an idealized scenario with isolated signal sources embedded in random noise. Real EEG data presents a more complex challenge where multiple interacting neural sources are simultaneously active, creating overlapping patterns of genuine dependencies. While our simplified model showed little information loss during volume conduction from sources to electrodes, real data involves many concurrent source signals that may exhibit nonlinear interactions during volume conduction to the scalp, further complicating the detection and interpretation of higher-order relationships. Additionally, forward modelling demonstrated that even ideal third-order interactions become difficult to detect below moderate SNRs. This implies that weak third-order interactions detected at the scalp, such as those identified in the motor imagery dataset, may reflect considerably stronger source-level dependencies.

Several methodological extensions warrant consideration. Our analysis examined only interactions among channel triplets within the same frequency band. Cross-frequency analysis would be necessary to capture the full complexity of neural interactions, though this increases computational burden through the number of unique triplet combinations. Similarly, we examined only triplets binarized using the same method. Mixed binarization could reveal phenomena such as phase-amplitude coupling, where high-amplitude periods in a higher frequency preferentially occur at specific phases of a lower frequency -- an interaction pattern observed in human cortical sources \cite{canolty2006High}. This could be detected by, for example, combining two Power-binarized signals with one Diff-binarized signal, though again at the cost of increased combinations to evaluate. Regarding computational scalability, the present analysis with 17 channels was feasible but may become prohibitive for high-density arrays of 128 or 256 channels.

Temporal resolution could also be enhanced. We used non-overlapping 1-second windows when calculating probability vectors, providing only coarse temporal dynamics. Overlapping windows with frequency-adaptive sizes -- larger for lower frequencies, smaller for higher frequencies -- could improve temporal resolution. The ability to track information changes over time is a key advantage of this approach as the temporal dynamics of interactions may provide insight as to their role. If interactions persist over time, they may reflect ongoing information sharing required for neural computation. If higher-order interactions are instead only momentary, they may reflect a role in rapidly synchronizing or recruiting different cortical modules. Overall, this IG methodology is promising for characterizing higher-order interactions in neural data and, by moving beyond pairwise measures, has the potential to advance our understanding of interacting systems in the brain.

\appendix

\section{Appendix}\label{AppendixA}

\begin{figure}[ht]
	\centering
	\includegraphics{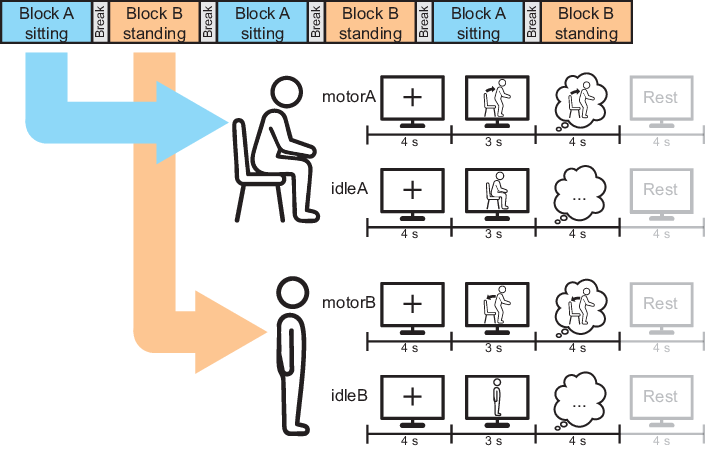}
	\caption{Design overview for the offline phase of the motor imagery experiment}
	\label{fig16}
\end{figure}

\subsection{Initial Filtering}\label{Initial Filtering}
All signals were initially filtered using a 0.5 Hz high-pass filter to remove low-frequency drifts. All signals were then notch filtered with cutoff frequencies of 58 Hz and 62 Hz to remove any 60 Hz electrical noise. All filtering steps used a finite impulse response (FIR) filter with the filter order determined automatically by the EEGLAB function.

\subsection{Artifact Removal}\label{Artifact Removal}
The data were cleaned using a multi-step process consisting of removing bad channels according to several criteria and correction non-stationary artifacts using Artifact Subspace Reconstruction (ASR) \cite{kothe2013BCILAB}. Channels were removed if any of the following default criteria were met: the signal was flat for more than five seconds, high frequency noise exceeded 4 SD, or the channel had a correlation coefficient with nearby channels $<0.8$. At this step, the mean number of rejected channels across participants was $0.5 \; [0,2]$. ASR was then performed using a threshold of 12 SD with a 0.5 s window, where data periods exceeding this threshold were interpolated. Following ASR, any channels previously rejected were interpolated using the spherical method in EEGLAB. The signals were then re-referenced to the average of all channels.
Independent component analysis was subsequently performed using the Extended Infomax algorithm to identify stationary artifacts. Independent components (ICs) were labeled using EEGLAB's IClabel tool: an automated IC classifier \cite{pion-tonachini2019ICLabel}. ICs are assigned probabilities of belonging to the following source categories: brain, muscle, eye, heart, line noise, channel noise, and other. ICs that were identified as non-brain activity with a probability of $\ge0.9$ were removed and the signals reconstructed. No ICs met the criteria for rejection during this step.

\backmatter

\bmhead{Acknowledgments}
This research was enabled in part by computing resources provided by the Prairies DRI Group and by the Digital Research Alliance of Canada (alliancecan.ca).

\bmhead{Author Contributions}
\textbf{Eric Albers:} Conceptualization, methodology, software, validation, formal analysis, writing -- original draft, visualization. \textbf{Paul Marriott:} Conceptualization, writing -- review and editing. \textbf{Masami Tatsuno:} Conceptualization, resources, writing -- review and editing, supervision, funding acquisition.

\bmhead{Funding}
This work was supported by NSERC Discovery Grant 2020-06342 (MT) and NSERC-Alberta Innovates Grant 232403195 (MT).

\bmhead{Availability of data and materials}
The EEG data that support the findings of this study are openly available from OpenNeuro at \url{https://doi.org/10.18112/openneuro.ds005342.v1.0.2} (cited in text). The code for EEG preprocessing and data analysis is available at \url{https://github.com/eric-albers/Using-Information-Geometry-to-Characterize-Higher-Order-Interactions-in-EEG}.

\section*{Declarations}

\bmhead{Conflict of interest}
The authors declare no conflict of interest. The authors have no relevant financial or non-financial interests to disclose.

\bibliography{bibliography}


\begin{thebibliography}{43}
\ifx \bisbn   \undefined \def \bisbn  #1{ISBN #1}\fi
\ifx \binits  \undefined \def \binits#1{#1}\fi
\ifx \bauthor  \undefined \def \bauthor#1{#1}\fi
\ifx \batitle  \undefined \def \batitle#1{#1}\fi
\ifx \bjtitle  \undefined \def \bjtitle#1{#1}\fi
\ifx \bvolume  \undefined \def \bvolume#1{\textbf{#1}}\fi
\ifx \byear  \undefined \def \byear#1{#1}\fi
\ifx \bissue  \undefined \def \bissue#1{#1}\fi
\ifx \bfpage  \undefined \def \bfpage#1{#1}\fi
\ifx \blpage  \undefined \def \blpage #1{#1}\fi
\ifx \burl  \undefined \def \burl#1{\textsf{#1}}\fi
\ifx \doiurl  \undefined \def \doiurl#1{\url{https://doi.org/#1}}\fi
\ifx \betal  \undefined \def \betal{\textit{et al.}}\fi
\ifx \binstitute  \undefined \def \binstitute#1{#1}\fi
\ifx \binstitutionaled  \undefined \def \binstitutionaled#1{#1}\fi
\ifx \bctitle  \undefined \def \bctitle#1{#1}\fi
\ifx \beditor  \undefined \def \beditor#1{#1}\fi
\ifx \bpublisher  \undefined \def \bpublisher#1{#1}\fi
\ifx \bbtitle  \undefined \def \bbtitle#1{#1}\fi
\ifx \bedition  \undefined \def \bedition#1{#1}\fi
\ifx \bseriesno  \undefined \def \bseriesno#1{#1}\fi
\ifx \blocation  \undefined \def \blocation#1{#1}\fi
\ifx \bsertitle  \undefined \def \bsertitle#1{#1}\fi
\ifx \bsnm \undefined \def \bsnm#1{#1}\fi
\ifx \bsuffix \undefined \def \bsuffix#1{#1}\fi
\ifx \bparticle \undefined \def \bparticle#1{#1}\fi
\ifx \barticle \undefined \def \barticle#1{#1}\fi
\bibcommenthead
\ifx \bconfdate \undefined \def \bconfdate #1{#1}\fi
\ifx \botherref \undefined \def \botherref #1{#1}\fi
\ifx \url \undefined \def \url#1{\textsf{#1}}\fi
\ifx \bchapter \undefined \def \bchapter#1{#1}\fi
\ifx \bbook \undefined \def \bbook#1{#1}\fi
\ifx \bcomment \undefined \def \bcomment#1{#1}\fi
\ifx \oauthor \undefined \def \oauthor#1{#1}\fi
\ifx \citeauthoryear \undefined \def \citeauthoryear#1{#1}\fi
\ifx \endbibitem  \undefined \def \endbibitem {}\fi
\ifx \bconflocation  \undefined \def \bconflocation#1{#1}\fi
\ifx \arxivurl  \undefined \def \arxivurl#1{\textsf{#1}}\fi
\csname PreBibitemsHook\endcsname

\bibitem[\protect\citeauthoryear{Kass et~al.}{2014}]{kass2014Analysis}
\begin{bbook}
\bauthor{\bsnm{Kass}, \binits{R.E.}},
\bauthor{\bsnm{Eden}, \binits{U.T.}},
\bauthor{\bsnm{Brown}, \binits{E.N.}}:
\bbtitle{Analysis of {{Neural Data}}}.
\bsertitle{Springer {{Series}} in {{Statistics}}}.
\bpublisher{Springer},
\blocation{New York}
(\byear{2014}).
\doiurl{10.1007/978-1-4614-9602-1}
\end{bbook}
\endbibitem

\bibitem[\protect\citeauthoryear{Amari and Nagaoka}{2000}]{amari2000Methods}
\begin{bbook}
\bauthor{\bsnm{Amari}, \binits{S.}},
\bauthor{\bsnm{Nagaoka}, \binits{H.}}:
\bbtitle{Methods of {{Information Geometry}}}.
\bsertitle{Translations of {{Mathematical Monographs}}},
vol. \bseriesno{191}.
\bpublisher{American Mathematical Society},
\blocation{Providence}
(\byear{2000}).
\doiurl{10.1090/mmono/191}
\end{bbook}
\endbibitem

\bibitem[\protect\citeauthoryear{Amari}{2001}]{amari2001Information}
\begin{barticle}
\bauthor{\bsnm{Amari}, \binits{S.}}:
\batitle{Information geometry on hierarchy of probability distributions}.
\bjtitle{IEEE Trans. Inf. Theory}
\bvolume{47}(\bissue{5}),
\bfpage{1701}--\blpage{1711}
(\byear{2001})
\doiurl{10.1109/18.930911}
\end{barticle}
\endbibitem

\bibitem[\protect\citeauthoryear{Nakahara and
  Amari}{2002}]{nakahara2002InformationGeometric}
\begin{barticle}
\bauthor{\bsnm{Nakahara}, \binits{H.}},
\bauthor{\bsnm{Amari}, \binits{S.}}:
\batitle{Information-geometric measure for neural spikes}.
\bjtitle{Neural Comput.}
\bvolume{14}(\bissue{10}),
\bfpage{2269}--\blpage{2316}
(\byear{2002})
\doiurl{10.1162/08997660260293238}
\end{barticle}
\endbibitem

\bibitem[\protect\citeauthoryear{Tatsuno
  et~al.}{2009}]{tatsuno2009InformationGeometric}
\begin{barticle}
\bauthor{\bsnm{Tatsuno}, \binits{M.}},
\bauthor{\bsnm{Fellous}, \binits{J.-M.}},
\bauthor{\bsnm{Amari}, \binits{S.}}:
\batitle{Information-geometric measures as robust estimators of connection
  strengths and external inputs}.
\bjtitle{Neural Comput.}
\bvolume{21}(\bissue{8}),
\bfpage{2309}--\blpage{2335}
(\byear{2009})
\doiurl{10.1162/neco.2009.04-08-748}
\end{barticle}
\endbibitem

\bibitem[\protect\citeauthoryear{Nie and
  Tatsuno}{2012}]{nie2012InformationGeometric}
\begin{barticle}
\bauthor{\bsnm{Nie}, \binits{Y.}},
\bauthor{\bsnm{Tatsuno}, \binits{M.}}:
\batitle{Information-geometric measures for estimation of connection weight
  under correlated inputs}.
\bjtitle{Neural Comput.}
\bvolume{24}(\bissue{12}),
\bfpage{3213}--\blpage{3245}
(\byear{2012})
\doiurl{10.1162/NECO_a_00367}
\end{barticle}
\endbibitem

\bibitem[\protect\citeauthoryear{Iwasaki et~al.}{2018}]{iwasaki2018Estimation}
\begin{barticle}
\bauthor{\bsnm{Iwasaki}, \binits{T.}},
\bauthor{\bsnm{Hino}, \binits{H.}},
\bauthor{\bsnm{Tatsuno}, \binits{M.}},
\bauthor{\bsnm{Akaho}, \binits{S.}},
\bauthor{\bsnm{Murata}, \binits{N.}}:
\batitle{Estimation of neural connections from partially observed neural
  spikes}.
\bjtitle{Neural Netw.}
\bvolume{108},
\bfpage{172}--\blpage{191}
(\byear{2018})
\doiurl{10.1016/j.neunet.2018.07.019}
\end{barticle}
\endbibitem

\bibitem[\protect\citeauthoryear{Kass et~al.}{2018}]{kass2018Computational}
\begin{barticle}
\bauthor{\bsnm{Kass}, \binits{R.E.}},
\bauthor{\bsnm{Amari}, \binits{S.}},
\bauthor{\bsnm{Arai}, \binits{K.}},
\bauthor{\bsnm{Brown}, \binits{E.N.}},
\bauthor{\bsnm{Diekman}, \binits{C.O.}},
\bauthor{\bsnm{Diesmann}, \binits{M.}},
\bauthor{\bsnm{Doiron}, \binits{B.}},
\bauthor{\bsnm{Eden}, \binits{U.T.}},
\bauthor{\bsnm{Fairhall}, \binits{A.L.}},
\bauthor{\bsnm{Fiddyment}, \binits{G.M.}},
\bauthor{\bsnm{Fukai}, \binits{T.}},
\bauthor{\bsnm{Gr{\"u}n}, \binits{S.}},
\bauthor{\bsnm{Harrison}, \binits{M.T.}},
\bauthor{\bsnm{Helias}, \binits{M.}},
\bauthor{\bsnm{Nakahara}, \binits{H.}},
\bauthor{\bsnm{Teramae}, \binits{J.}},
\bauthor{\bsnm{Thomas}, \binits{P.J.}},
\bauthor{\bsnm{Reimers}, \binits{M.}},
\bauthor{\bsnm{Rodu}, \binits{J.}},
\bauthor{\bsnm{Rotstein}, \binits{H.G.}},
\bauthor{\bsnm{{Shea-Brown}}, \binits{E.}},
\bauthor{\bsnm{Shimazaki}, \binits{H.}},
\bauthor{\bsnm{Shinomoto}, \binits{S.}},
\bauthor{\bsnm{Yu}, \binits{B.M.}},
\bauthor{\bsnm{Kramer}, \binits{M.A.}}:
\batitle{Computational neuroscience: {{Mathematical}} and statistical
  perspectives}.
\bjtitle{Annu. Rev. Stat. Appl.}
\bvolume{5}(\bissue{1}),
\bfpage{183}--\blpage{214}
(\byear{2018})
\doiurl{10.1146/annurev-statistics-041715-033733}
\end{barticle}
\endbibitem

\bibitem[\protect\citeauthoryear{Crosser and
  Brinkman}{2024}]{crosser2024Applications}
\begin{barticle}
\bauthor{\bsnm{Crosser}, \binits{J.T.}},
\bauthor{\bsnm{Brinkman}, \binits{B.A.W.}}:
\batitle{Applications of information geometry to spiking neural network
  activity}.
\bjtitle{Phys. Rev. E}
\bvolume{109},
\bfpage{024302}
(\byear{2024})
\doiurl{10.1103/PhysRevE.109.024302}
\end{barticle}
\endbibitem

\bibitem[\protect\citeauthoryear{Kandel et~al.}{2021}]{kandel2021Principles}
\begin{bbook}
\beditor{\bsnm{Kandel}, \binits{E.R.}},
\beditor{\bsnm{Koester}, \binits{J.D.}},
\beditor{\bsnm{Mack}, \binits{S.H.}},
\beditor{\bsnm{Siegelbaum}, \binits{S.A.}} (eds.):
\bbtitle{Principles of {{Neural Science}}},
\bedition{6th} edn.
\bpublisher{McGraw Hill},
\blocation{New York}
(\byear{2021})
\end{bbook}
\endbibitem

\bibitem[\protect\citeauthoryear{Schomer and {Lopes da
  Silva}}{2011}]{schomer2011Niedermeyers}
\begin{bbook}
\beditor{\bsnm{Schomer}, \binits{D.L.}},
\beditor{\bsnm{{Lopes da Silva}}, \binits{F.H.}} (eds.):
\bbtitle{Niedermeyer's {{Electroencephalography}}: {{Basic Principles}},
  {{Clinical Applications}}, and {{Related Fields}}},
\bedition{6th} edn.
\bpublisher{Lippincott Williams \& Wilkins},
\blocation{Philadelphia}
(\byear{2011})
\end{bbook}
\endbibitem

\bibitem[\protect\citeauthoryear{Bressler and
  Menon}{2010}]{bressler2010Largescale}
\begin{barticle}
\bauthor{\bsnm{Bressler}, \binits{S.L.}},
\bauthor{\bsnm{Menon}, \binits{V.}}:
\batitle{Large-scale brain networks in cognition: {{Emerging}} methods and
  principles}.
\bjtitle{Trends Cogn. Sci.}
\bvolume{14}(\bissue{6}),
\bfpage{277}--\blpage{290}
(\byear{2010})
\doiurl{10.1016/j.tics.2010.04.004}
\end{barticle}
\endbibitem

\bibitem[\protect\citeauthoryear{Sporns et~al.}{2005}]{sporns2005Human}
\begin{barticle}
\bauthor{\bsnm{Sporns}, \binits{O.}},
\bauthor{\bsnm{Tononi}, \binits{G.}},
\bauthor{\bsnm{K{\"o}tter}, \binits{R.}}:
\batitle{The human connectome: {{A}} structural description of the human
  brain}.
\bjtitle{PLoS Comput. Biol.}
\bvolume{1}(\bissue{4}),
\bfpage{42}
(\byear{2005})
\doiurl{10.1371/journal.pcbi.0010042}
\end{barticle}
\endbibitem

\bibitem[\protect\citeauthoryear{Elam et~al.}{2021}]{elam2021Human}
\begin{barticle}
\bauthor{\bsnm{Elam}, \binits{J.S.}},
\bauthor{\bsnm{Glasser}, \binits{M.F.}},
\bauthor{\bsnm{Harms}, \binits{M.P.}},
\bauthor{\bsnm{Sotiropoulos}, \binits{S.N.}},
\bauthor{\bsnm{Andersson}, \binits{J.L.R.}},
\bauthor{\bsnm{Burgess}, \binits{G.C.}},
\bauthor{\bsnm{Curtiss}, \binits{S.W.}},
\bauthor{\bsnm{Oostenveld}, \binits{R.}},
\bauthor{\bsnm{{Larson-Prior}}, \binits{L.J.}},
\bauthor{\bsnm{Schoffelen}, \binits{J.-M.}},
\bauthor{\bsnm{Hodge}, \binits{M.R.}},
\bauthor{\bsnm{Cler}, \binits{E.A.}},
\bauthor{\bsnm{Marcus}, \binits{D.M.}},
\bauthor{\bsnm{Barch}, \binits{D.M.}},
\bauthor{\bsnm{Yacoub}, \binits{E.}},
\bauthor{\bsnm{Smith}, \binits{S.M.}},
\bauthor{\bsnm{Ugurbil}, \binits{K.}},
\bauthor{\bsnm{Van~Essen}, \binits{D.C.}}:
\batitle{The {{Human Connectome Project}}: {{A}} retrospective}.
\bjtitle{NeuroImage}
\bvolume{244},
\bfpage{118543}
(\byear{2021})
\doiurl{10.1016/j.neuroimage.2021.118543}
\end{barticle}
\endbibitem

\bibitem[\protect\citeauthoryear{Friston}{1994}]{friston1994Functional}
\begin{barticle}
\bauthor{\bsnm{Friston}, \binits{K.J.}}:
\batitle{Functional and effective connectivity in neuroimaging: {{A}}
  synthesis}.
\bjtitle{Hum. Brain Mapp.}
\bvolume{2}(\bissue{1-2}),
\bfpage{56}--\blpage{78}
(\byear{1994})
\doiurl{10.1002/hbm.460020107}
\end{barticle}
\endbibitem

\bibitem[\protect\citeauthoryear{Bullmore and
  Sporns}{2009}]{bullmore2009Complex}
\begin{barticle}
\bauthor{\bsnm{Bullmore}, \binits{E.}},
\bauthor{\bsnm{Sporns}, \binits{O.}}:
\batitle{Complex brain networks: {{Graph}} theoretical analysis of structural
  and functional systems}.
\bjtitle{Nat. Rev. Neurosci.}
\bvolume{10},
\bfpage{186}--\blpage{198}
(\byear{2009})
\doiurl{10.1038/nrn2575}
\end{barticle}
\endbibitem

\bibitem[\protect\citeauthoryear{Thomas~Yeo
  et~al.}{2011}]{thomasyeo2011Organization}
\begin{barticle}
\bauthor{\bsnm{Thomas~Yeo}, \binits{B.T.}},
\bauthor{\bsnm{Krienen}, \binits{F.M.}},
\bauthor{\bsnm{Sepulcre}, \binits{J.}},
\bauthor{\bsnm{Sabuncu}, \binits{M.R.}},
\bauthor{\bsnm{Lashkari}, \binits{D.}},
\bauthor{\bsnm{Hollinshead}, \binits{M.}},
\bauthor{\bsnm{Roffman}, \binits{J.L.}},
\bauthor{\bsnm{Smoller}, \binits{J.W.}},
\bauthor{\bsnm{Z{\"o}llei}, \binits{L.}},
\bauthor{\bsnm{Polimeni}, \binits{J.R.}},
\bauthor{\bsnm{Fischl}, \binits{B.}},
\bauthor{\bsnm{Liu}, \binits{H.}},
\bauthor{\bsnm{Buckner}, \binits{R.L.}}:
\batitle{The organization of the human cerebral cortex estimated by intrinsic
  functional connectivity}.
\bjtitle{J. Neurophysiol.}
\bvolume{106}(\bissue{3}),
\bfpage{1125}--\blpage{1165}
(\byear{2011})
\doiurl{10.1152/jn.00338.2011}
\end{barticle}
\endbibitem

\bibitem[\protect\citeauthoryear{Bassett and Sporns}{2017}]{bassett2017Network}
\begin{barticle}
\bauthor{\bsnm{Bassett}, \binits{D.S.}},
\bauthor{\bsnm{Sporns}, \binits{O.}}:
\batitle{Network neuroscience}.
\bjtitle{Nat. Neurosci.}
\bvolume{20}(\bissue{3}),
\bfpage{353}--\blpage{364}
(\byear{2017})
\doiurl{10.1038/nn.4502}
\end{barticle}
\endbibitem

\bibitem[\protect\citeauthoryear{Damoiseaux}{2017}]{damoiseaux2017Effects}
\begin{barticle}
\bauthor{\bsnm{Damoiseaux}, \binits{J.S.}}:
\batitle{Effects of aging on functional and structural brain connectivity}.
\bjtitle{NeuroImage}
\bvolume{160},
\bfpage{32}--\blpage{40}
(\byear{2017})
\doiurl{10.1016/j.neuroimage.2017.01.077}
\end{barticle}
\endbibitem

\bibitem[\protect\citeauthoryear{Seeley
  et~al.}{2009}]{seeley2009Neurodegenerative}
\begin{barticle}
\bauthor{\bsnm{Seeley}, \binits{W.W.}},
\bauthor{\bsnm{Crawford}, \binits{R.K.}},
\bauthor{\bsnm{Zhou}, \binits{J.}},
\bauthor{\bsnm{Miller}, \binits{B.L.}},
\bauthor{\bsnm{Greicius}, \binits{M.D.}}:
\batitle{Neurodegenerative diseases target large-scale human brain networks}.
\bjtitle{Neuron}
\bvolume{62}(\bissue{1}),
\bfpage{42}--\blpage{52}
(\byear{2009})
\doiurl{10.1016/j.neuron.2009.03.024}
\end{barticle}
\endbibitem

\bibitem[\protect\citeauthoryear{Battiston
  et~al.}{2020}]{battiston2020Networks}
\begin{barticle}
\bauthor{\bsnm{Battiston}, \binits{F.}},
\bauthor{\bsnm{Cencetti}, \binits{G.}},
\bauthor{\bsnm{Iacopini}, \binits{I.}},
\bauthor{\bsnm{Latora}, \binits{V.}},
\bauthor{\bsnm{Lucas}, \binits{M.}},
\bauthor{\bsnm{Patania}, \binits{A.}},
\bauthor{\bsnm{Young}, \binits{J.-G.}},
\bauthor{\bsnm{Petri}, \binits{G.}}:
\batitle{Networks beyond pairwise interactions: {{Structure}} and dynamics}.
\bjtitle{Phys. Rep.}
\bvolume{874},
\bfpage{1}--\blpage{92}
(\byear{2020})
\doiurl{10.1016/j.physrep.2020.05.004}
\end{barticle}
\endbibitem

\bibitem[\protect\citeauthoryear{Battiston et~al.}{2021}]{battiston2021Physics}
\begin{barticle}
\bauthor{\bsnm{Battiston}, \binits{F.}},
\bauthor{\bsnm{Amico}, \binits{E.}},
\bauthor{\bsnm{Barrat}, \binits{A.}},
\bauthor{\bsnm{Bianconi}, \binits{G.}},
\bauthor{\bsnm{{Ferraz de Arruda}}, \binits{G.}},
\bauthor{\bsnm{Franceschiello}, \binits{B.}},
\bauthor{\bsnm{Iacopini}, \binits{I.}},
\bauthor{\bsnm{K{\'e}fi}, \binits{S.}},
\bauthor{\bsnm{Latora}, \binits{V.}},
\bauthor{\bsnm{Moreno}, \binits{Y.}},
\bauthor{\bsnm{Murray}, \binits{M.M.}},
\bauthor{\bsnm{Peixoto}, \binits{T.P.}},
\bauthor{\bsnm{Vaccarino}, \binits{F.}},
\bauthor{\bsnm{Petri}, \binits{G.}}:
\batitle{The physics of higher-order interactions in complex systems}.
\bjtitle{Nat. Phys.}
\bvolume{17},
\bfpage{1093}--\blpage{1098}
(\byear{2021})
\doiurl{10.1038/s41567-021-01371-4}
\end{barticle}
\endbibitem

\bibitem[\protect\citeauthoryear{{Barndorff-Nielsen}}{2014}]{barndorff-nielsen2014Information}
\begin{bbook}
\bauthor{\bsnm{{Barndorff-Nielsen}}, \binits{O.}}:
\bbtitle{Information and {{Exponential Families}}: {{In Statistical Theory}}}.
\bsertitle{Wiley {{Series}} in {{Probability}} and {{Statistics}}}.
\bpublisher{John Wiley \& Sons},
\blocation{Chichester}
(\byear{2014}).
\doiurl{10.1002/9781118857281}
\end{bbook}
\endbibitem

\bibitem[\protect\citeauthoryear{Brown}{1986}]{brown1986Fundamentals}
\begin{bbook}
\bauthor{\bsnm{Brown}, \binits{L.D.}}:
\bbtitle{Fundamentals of {{Statistical Exponential Families}} with
  {{Applications}} in {{Statistical Decision Theory}}}.
\bsertitle{Lecture {{Notes}} - {{Monograph Series}}},
vol. \bseriesno{9}.
\bpublisher{Institute of Mathematical Statistics},
\blocation{Hayward}
(\byear{1986}).
\doiurl{10.1214/lnms/1215466757}
\end{bbook}
\endbibitem

\bibitem[\protect\citeauthoryear{Csisz{\'a}r and Mat{\'u}{\v
  s}}{2005}]{csiszar2005Closures}
\begin{barticle}
\bauthor{\bsnm{Csisz{\'a}r}, \binits{I.}},
\bauthor{\bsnm{Mat{\'u}{\v s}}, \binits{F.}}:
\batitle{Closures of exponential families}.
\bjtitle{Ann. Probab.}
\bvolume{33}(\bissue{2}),
\bfpage{582}--\blpage{600}
(\byear{2005})
\doiurl{10.1214/009117904000000766}
\end{barticle}
\endbibitem

\bibitem[\protect\citeauthoryear{{Anaya-Izquierdo}
  et~al.}{2014}]{anaya-izquierdo2014When}
\begin{barticle}
\bauthor{\bsnm{{Anaya-Izquierdo}}, \binits{K.}},
\bauthor{\bsnm{Critchley}, \binits{F.}},
\bauthor{\bsnm{Marriott}, \binits{P.}}:
\batitle{When are first-order asymptotics adequate? {{A}} diagnostic}.
\bjtitle{Stat}
\bvolume{3}(\bissue{1}),
\bfpage{17}--\blpage{22}
(\byear{2014})
\doiurl{10.1002/sta4.40}
\end{barticle}
\endbibitem

\bibitem[\protect\citeauthoryear{Yates}{1934}]{yates1934Contingency}
\begin{barticle}
\bauthor{\bsnm{Yates}, \binits{F.}}:
\batitle{Contingency tables involving small numbers and the {$\chi$}2 test}.
\bjtitle{Suppl. J. R. Stat. Soc.}
\bvolume{1}(\bissue{2}),
\bfpage{217}--\blpage{235}
(\byear{1934})
\doiurl{10.2307/2983604}
\end{barticle}
\endbibitem

\bibitem[\protect\citeauthoryear{Haber}{1982}]{haber1982Continuity}
\begin{barticle}
\bauthor{\bsnm{Haber}, \binits{M.}}:
\batitle{The continuity correction and statistical testing}.
\bjtitle{Int. Stat. Rev.}
\bvolume{50}(\bissue{2}),
\bfpage{135}--\blpage{144}
(\byear{1982})
\doiurl{10.2307/1402597}
{\href{https://arxiv.org/abs/1402597}{{1402597}}}
\end{barticle}
\endbibitem

\bibitem[\protect\citeauthoryear{Brown et~al.}{2001}]{brown2001Interval}
\begin{barticle}
\bauthor{\bsnm{Brown}, \binits{L.D.}},
\bauthor{\bsnm{Cai}, \binits{T.T.}},
\bauthor{\bsnm{DasGupta}, \binits{A.}}:
\batitle{Interval estimation for a binomial proportion}.
\bjtitle{Stat. Sci.}
\bvolume{16}(\bissue{2}),
\bfpage{101}--\blpage{133}
(\byear{2001})
\doiurl{10.1214/ss/1009213286}
\end{barticle}
\endbibitem

\bibitem[\protect\citeauthoryear{Emura and Liao}{2018}]{emura2018Critical}
\begin{barticle}
\bauthor{\bsnm{Emura}, \binits{T.}},
\bauthor{\bsnm{Liao}, \binits{Y.-T.}}:
\batitle{Critical review and comparison of continuity correction methods:
  {{The}} normal approximation to the binomial distribution}.
\bjtitle{Commun. Stat. Simul. Comput.}
\bvolume{47}(\bissue{8}),
\bfpage{2266}--\blpage{2285}
(\byear{2018})
\doiurl{10.1080/03610918.2017.1341527}
\end{barticle}
\endbibitem

\bibitem[\protect\citeauthoryear{{Triana-Guzman}
  et~al.}{2024}]{triana-guzman2024EEG}
\begin{botherref}
\oauthor{\bsnm{{Triana-Guzman}}, \binits{N.}},
\oauthor{\bsnm{{Orjuela-Ca{\~n}on}}, \binits{A.D.}},
\oauthor{\bsnm{Jutinico}, \binits{A.L.}},
\oauthor{\bsnm{{Mendoza-Montoya}}, \binits{O.}},
\oauthor{\bsnm{Antelis}, \binits{J.M.}}:
{{EEG}} Data Offline and Online during Motor Imagery for Standing and Sitting.
OpenNeuro
(2024).
\doiurl{10.18112/OPENNEURO.DS005342.V1.0.3}
\end{botherref}
\endbibitem

\bibitem[\protect\citeauthoryear{{Triana-Guzman}
  et~al.}{2022}]{triana-guzman2022Decoding}
\begin{barticle}
\bauthor{\bsnm{{Triana-Guzman}}, \binits{N.}},
\bauthor{\bsnm{{Orjuela-Ca{\~n}on}}, \binits{A.D.}},
\bauthor{\bsnm{Jutinico}, \binits{A.L.}},
\bauthor{\bsnm{{Mendoza-Montoya}}, \binits{O.}},
\bauthor{\bsnm{Antelis}, \binits{J.M.}}:
\batitle{Decoding {{EEG}} rhythms offline and online during motor imagery for
  standing and sitting based on a brain-computer interface}.
\bjtitle{Front. Neuroinformatics}
\bvolume{16},
\bfpage{961089}
(\byear{2022})
\doiurl{10.3389/fninf.2022.961089}
\end{barticle}
\endbibitem

\bibitem[\protect\citeauthoryear{Sharbrough
  et~al.}{1991}]{sharbrough1991American}
\begin{barticle}
\bauthor{\bsnm{Sharbrough}, \binits{F.}},
\bauthor{\bsnm{Chatrian}, \binits{G.-E.}},
\bauthor{\bsnm{Lesser}, \binits{R.P.}},
\bauthor{\bsnm{L{\"u}ders}, \binits{H.}},
\bauthor{\bsnm{Nuwer}, \binits{M.}},
\bauthor{\bsnm{Picton}, \binits{T.W.}}:
\batitle{American {{Electroencephalographic Society}} guidelines for standard
  electrode position nomenclature}.
\bjtitle{J. Clin. Neurophysiol.}
\bvolume{8}(\bissue{2}),
\bfpage{200}--\blpage{202}
(\byear{1991})
\doiurl{10.1097/00004691-199104000-00007}
\end{barticle}
\endbibitem

\bibitem[\protect\citeauthoryear{Delorme and Makeig}{2004}]{delorme2004EEGLAB}
\begin{barticle}
\bauthor{\bsnm{Delorme}, \binits{A.}},
\bauthor{\bsnm{Makeig}, \binits{S.}}:
\batitle{{{EEGLAB}}: {{An}} open source toolbox for analysis of single-trial
  {{EEG}} dynamics including independent component analysis}.
\bjtitle{J. Neurosci. Methods}
\bvolume{134}(\bissue{1}),
\bfpage{9}--\blpage{21}
(\byear{2004})
\doiurl{10.1016/j.jneumeth.2003.10.009}
\end{barticle}
\endbibitem

\bibitem[\protect\citeauthoryear{Schreiber and
  Schmitz}{1996}]{schreiber1996Improved}
\begin{barticle}
\bauthor{\bsnm{Schreiber}, \binits{T.}},
\bauthor{\bsnm{Schmitz}, \binits{A.}}:
\batitle{Improved surrogate data for nonlinearity tests}.
\bjtitle{Phys. Rev. Lett.}
\bvolume{77},
\bfpage{635}--\blpage{638}
(\byear{1996})
\doiurl{10.1103/PhysRevLett.77.635}
\end{barticle}
\endbibitem

\bibitem[\protect\citeauthoryear{Oostenveld
  et~al.}{2011}]{oostenveld2011FieldTrip}
\begin{barticle}
\bauthor{\bsnm{Oostenveld}, \binits{R.}},
\bauthor{\bsnm{Fries}, \binits{P.}},
\bauthor{\bsnm{Maris}, \binits{E.}},
\bauthor{\bsnm{Schoffelen}, \binits{J.-M.}}:
\batitle{{{FieldTrip}}: {{Open}} source software for advanced analysis of
  {{MEG}}, {{EEG}}, and invasive electrophysiological data}.
\bjtitle{Comput. Intell. Neurosci.}
\bvolume{2011},
\bfpage{156869}
(\byear{2011})
\doiurl{10.1155/2011/156869}
\end{barticle}
\endbibitem

\bibitem[\protect\citeauthoryear{Oostendorp and {van
  Oosterom}}{1991}]{oostendorp1991Potential}
\begin{barticle}
\bauthor{\bsnm{Oostendorp}, \binits{T.}},
\bauthor{\bsnm{{van Oosterom}}, \binits{A.}}:
\batitle{The potential distribution generated by surface electrodes in
  inhomogeneous volume conductors of arbitrary shape}.
\bjtitle{IEEE Trans. Biomed. Eng.}
\bvolume{38}(\bissue{5}),
\bfpage{409}--\blpage{417}
(\byear{1991})
\doiurl{10.1109/10.81559}
\end{barticle}
\endbibitem

\bibitem[\protect\citeauthoryear{Oostenveld et~al.}{2003}]{oostenveld2003Brain}
\begin{barticle}
\bauthor{\bsnm{Oostenveld}, \binits{R.}},
\bauthor{\bsnm{Stegeman}, \binits{D.F.}},
\bauthor{\bsnm{Praamstra}, \binits{P.}},
\bauthor{\bsnm{{van Oosterom}}, \binits{A.}}:
\batitle{Brain symmetry and topographic analysis of lateralized event-related
  potentials}.
\bjtitle{Clin. Neurophysiol.}
\bvolume{114}(\bissue{7}),
\bfpage{1194}--\blpage{1202}
(\byear{2003})
\doiurl{10.1016/S1388-2457(03)00059-2}
\end{barticle}
\endbibitem

\bibitem[\protect\citeauthoryear{Holmes et~al.}{1998}]{holmes1998Enhancement}
\begin{barticle}
\bauthor{\bsnm{Holmes}, \binits{C.J.}},
\bauthor{\bsnm{Hoge}, \binits{R.}},
\bauthor{\bsnm{Collins}, \binits{L.}},
\bauthor{\bsnm{Woods}, \binits{R.}},
\bauthor{\bsnm{Toga}, \binits{A.W.}},
\bauthor{\bsnm{Evans}, \binits{A.C.}}:
\batitle{Enhancement of {{MR}} images using registration for signal averaging}.
\bjtitle{J. Comput. Assist. Tomogr.}
\bvolume{22}(\bissue{2}),
\bfpage{324}--\blpage{333}
(\byear{1998})
\doiurl{10.1097/00004728-199803000-00032}
\end{barticle}
\endbibitem

\bibitem[\protect\citeauthoryear{Guillot et~al.}{2009}]{guillot2009Brain}
\begin{barticle}
\bauthor{\bsnm{Guillot}, \binits{A.}},
\bauthor{\bsnm{Collet}, \binits{C.}},
\bauthor{\bsnm{Nguyen}, \binits{V.A.}},
\bauthor{\bsnm{Malouin}, \binits{F.}},
\bauthor{\bsnm{Richards}, \binits{C.}},
\bauthor{\bsnm{Doyon}, \binits{J.}}:
\batitle{Brain activity during visual versus kinesthetic imagery: {{An fMRI}}
  study}.
\bjtitle{Hum. Brain Mapp.}
\bvolume{30}(\bissue{7}),
\bfpage{2157}--\blpage{2172}
(\byear{2009})
\doiurl{10.1002/hbm.20658}
\end{barticle}
\endbibitem

\bibitem[\protect\citeauthoryear{Canolty et~al.}{2006}]{canolty2006High}
\begin{barticle}
\bauthor{\bsnm{Canolty}, \binits{R.T.}},
\bauthor{\bsnm{Edwards}, \binits{E.}},
\bauthor{\bsnm{Dalal}, \binits{S.S.}},
\bauthor{\bsnm{Soltani}, \binits{M.}},
\bauthor{\bsnm{Nagarajan}, \binits{S.S.}},
\bauthor{\bsnm{Kirsch}, \binits{H.E.}},
\bauthor{\bsnm{Berger}, \binits{M.S.}},
\bauthor{\bsnm{Barbaro}, \binits{N.M.}},
\bauthor{\bsnm{Knight}, \binits{R.T.}}:
\batitle{High gamma power is phase-locked to theta oscillations in human
  neocortex}.
\bjtitle{Science}
\bvolume{313}(\bissue{5793}),
\bfpage{1626}--\blpage{1628}
(\byear{2006})
\doiurl{10.1126/science.1128115}
\end{barticle}
\endbibitem

\bibitem[\protect\citeauthoryear{Kothe and Makeig}{2013}]{kothe2013BCILAB}
\begin{barticle}
\bauthor{\bsnm{Kothe}, \binits{C.A.}},
\bauthor{\bsnm{Makeig}, \binits{S.}}:
\batitle{{{BCILAB}}: {{A}} platform for brain--computer interface development}.
\bjtitle{J. Neural Eng.}
\bvolume{10}(\bissue{5}),
\bfpage{056014}
(\byear{2013})
\doiurl{10.1088/1741-2560/10/5/056014}
\end{barticle}
\endbibitem

\bibitem[\protect\citeauthoryear{{Pion-Tonachini}
  et~al.}{2019}]{pion-tonachini2019ICLabel}
\begin{barticle}
\bauthor{\bsnm{{Pion-Tonachini}}, \binits{L.}},
\bauthor{\bsnm{{Kreutz-Delgado}}, \binits{K.}},
\bauthor{\bsnm{Makeig}, \binits{S.}}:
\batitle{{{ICLabel}}: {{An}} automated electroencephalographic independent
  component classifier, dataset, and website}.
\bjtitle{NeuroImage}
\bvolume{198},
\bfpage{181}--\blpage{197}
(\byear{2019})
\doiurl{10.1016/j.neuroimage.2019.05.026}
\end{barticle}
\endbibitem

\end{thebibliography}

\end{document}